\documentclass{aastex}
\usepackage{graphicx, subfigure, multirow}

\begin{document}
 
\title{A Multi-instrument Analysis of Sunspot Umbrae}

\author{F. T. Watson, M. J. Penn and W. Livingston}
\affil{National Solar Observatory, 950 N. Cherry Ave., Tucson, AZ 85719, USA}
\email{fwatson@nso.edu}

\begin{abstract}
The recent solar minimum and rise phase of solar cycle 24 have been unlike any
period since the early 1900s. This article examines some of the properties of
sunspot umbrae over the last 17 years with three different instruments on the ground and in space: MDI, HMI and BABO. The distribution of magnetic fields and their evolution over time is shown and reveals that the field distribution in cycle 24 is
fundamentally different from that in cycle 23. The annual average umbral
magnetic field is then examined for the 17 year observation period and shows a small decrease of 375~Gauss in sunspot magnetic fields over the period 1996--2013, but the mean intensity of sunspot umbrae does not vary significantly over this time. A possible
issue with sample sizes in a previous study is then explored to explain
disagreements in data from two of the source instruments. All three instruments show that the relationship between umbral magnetic fields and umbral intensity agrees with past studies in that the umbral intensity decreases as the field strength increases. This apparent contradiction can be explained by the range of magnetic field values measured for a given umbral intensity being larger than the measured 375~G change in umbral field strength over time. 
\end{abstract}


\section{Introduction}\label{sect:intro}
The study of sunspots is of great importance in modern solar physics as they are
one of the primary manifestations of the solar magnetic field and their
properties are used as proxies in many different areas of solar and terrestrial research. The
evolution of sunspot magnetic fields throughout a solar cycle has been
an area of active study over the last few years. \cite{2010SoPh..262...19S} used
Kitt Peak Vacuum Telescope data to show a fairly constant maximum umbral field strength between 1993
and 2003. \cite{2011watson} used data from MDI to show that the average maximum
umbral field strength measured on the Sun varied as a function of the solar
cycle and this result was confirmed by \cite{2011ApJ...742L..36P} using data
from seven former USSR observatories. \cite{2012rezaei} used the Tenerife
Infrared Polarimeter to measure umbral magnetic fields and found a very small decrease in
the average maximum umbral field strength but concluded that the field strengths are dominated by cyclic variations. \cite{2012ApJ...757L...8L} have
also seen a decrease in average maximum umbral field strengths of around 2600~G
in 1999 to around 2100~G in 2012. Most recently, \cite{2013SoPh..tmp...16P} used
data from Mt. Wilson Observatory and showed that taking the strongest magnetic
field measurement on a weekly basis gave a solar cycle dependent trend in the
average maximum umbral field strengths.

The three studies that showed a solar cycle dependance of the umbral magnetic
field (\cite{2011watson}, \cite{2011ApJ...742L..36P} and
\cite{2013SoPh..tmp...16P}) all have one thing in common. They do not measure
all of the sunspot umbrae and only record the strongest umbra observed on any given day. By measuring only the strongest
sunspot on the disk in any given time period, a cyclic pattern should be
expected as preference is given to the strongest sunspot on the disk. There are
many more weaker sunspots and so by measuring the strongest field in all sunspot
umbrae, more information about the global magnetic field properties can be
obtained. This work uses an automated sunspot detection algorithm (see \cite{2009SoPh..260....5W} for details) to find
all sunspots present in datasets from space-based observatories and create a
catalog. This catalog allows a statistical analysis of various properties
using more sunspots than has been possible in previous works.

Section~\ref{sect:data} describes the data used in the analysis of sunspots and
introduces the catalogs. Then, Section~\ref{sect:B_vs_I} compares the
intensity of sunspot umbrae to the magnetic field measured in the same location.
In Section~\ref{sect:field_dist} the distribution of magnetic field strengths in
sunspot umbrae is analysed and Section~\ref{sect:temporal} continues this by
looking at the umbral magnetic fields and intensities as they change over time.
Finally, a summary with conclusions is delivered in Section~\ref{sect:conclusions}.

\section{Data}\label{sect:data}
This article draws data from different sources, both space and
ground-based.

\subsection{Space-based}\label{sect:data_space}
Two instruments are used to obtain continuum and magnetogram data for sunspots.
The first of these is the Michelson Doppler Imager (MDI,
\cite{1995SoPh..162..129S}) which was launched in 1995 onboard the Solar and
Heliospheric Observatory (SOHO). MDI was capable of producing intensitygrams at
a 6 hour cadence and magnetograms at a 96 minute cadence with 4.0 arc-second
spatial resolution and did so until 2010 when a similar instrument was launched
on the Solar Dynamics Observatory (SDO) spacecraft.

The Helioseismic and Magnetic Imager (HMI, \cite{2012SoPh..275..207S}) was
launched onboard SDO in 2010 and is capable of the same measurements as MDI but
at an improved spatial and temporal resolution as well as being able to more
accurately measure the magnetic field on the Sun. Intensitygrams and
magnetograms are available with 45 second cadence and at 1.0 arc-second spatial
resolution.

\subsection{Ground-based}\label{sect:data_ground}
The ground-based observations used in this study are the collection of infrared
sunspot umbrae measurements taken with the Baboquivari detector (BABO) at the National Solar Observatory McMath-Pierce Solar Telescope from 1999-2013. The spectrum of the Fe I line at 1564.8 nm is recorded and the Zeeman splitting of the line is directly measured to give the magnetic field strength in the darkest part of sunspot umbrae. The wavelength splitting is
completely independent of viewing angle and is fully resolved for magnetic fields greater than around 1100 Gauss. This is described in more detail in \cite{2006ApJ...649L..45P} and  \cite{2012ApJ...757L...8L}.

\subsection{Comparison}\label{sect:comparison}
Both the space and ground-based datasets come with advantages and disadvantages
for this type of study. The MDI and HMI data benefit from near-complete temporal
coverage during their missions. As such, there is data available for almost
every day in the period 1995-2013 (excluding some time in 1998 when SOHO was
essentially lost for four months). In comparison, the infrared measurements at
the McMath-Pierce Solar Telescope are less regular, depending on the amount of
observation time
allocated as well as weather conditions. For most of the period of interest, this amounts to 5-7 days per
month. As such, the space record is a more complete temporal picture of the sunspot
population.

The method of measuring the magnetic field used by MDI and HMI is to take a
number of points along the spectral line profile and fit them to a model. In
doing so, the line-of-sight magnetic field can be inferred. However, the
infrared measurements show the whole spectrum from 1564.4--1565.6~nm and allows the true continuum as well as the magnetic field at the source to be directly measured.

\textbf{A major disadvantage of the MDI measurements is that only the line-of-sight magnetic field strength is obtained. As we measure the darkest part of sunspot umbrae, we expect this to have the strongest magnetic field within the sunspot and correct the line-of-sight field assuming that the field is normal to the photosphere.}

The HMI magnetogram data used are Stokes profile inversions using the
Milne-Eddington model of stellar atmospheres, which are now one of the standard
data products produced by the HMI team. \textbf{Details of the method used are explained in detail in \cite{2007SoPh..240..177B} and \cite{2011SoPh..273..267B}}. As a result, we expect the HMI magnetic
fields to be more reliable than those from MDI where it is assumed that the
magnetic field in the darkest part of the sunspot umbra is normal to the
photosphere and a simple cosine correction is applied for line of sight effects.

\subsection{Catalogs}
Different methods are used to choose sunspots from each of the data sets. The
MDI and HMI data is analysed using the STARA code \citep{2009SoPh..260....5W}
which is an automated detection algorithm that allows sunspot properties to be
extracted from large datasets. The infrared catalog is comprised of the
visible sunspots on the disk when observing time was available, and as mentioned
previously, this is around 5-7 days per month. For all datasets, when data is available one
measurement per umbra is made per day. The MDI,
HMI and
infrared catalogs contain 17379, 9542 and 3949 entries respectively. It should
be noted that these are not all individual sunspots, but individual detections.
If a sunspot persists for 12 days, it will have 12 entries in each catalog,
assuming data was available.

\begin{figure}[ht!]
 \centering
 \subfigure[MDI data]{
            \label{fig:B_vs_I_MDI}
            \includegraphics[width=0.45\textwidth]{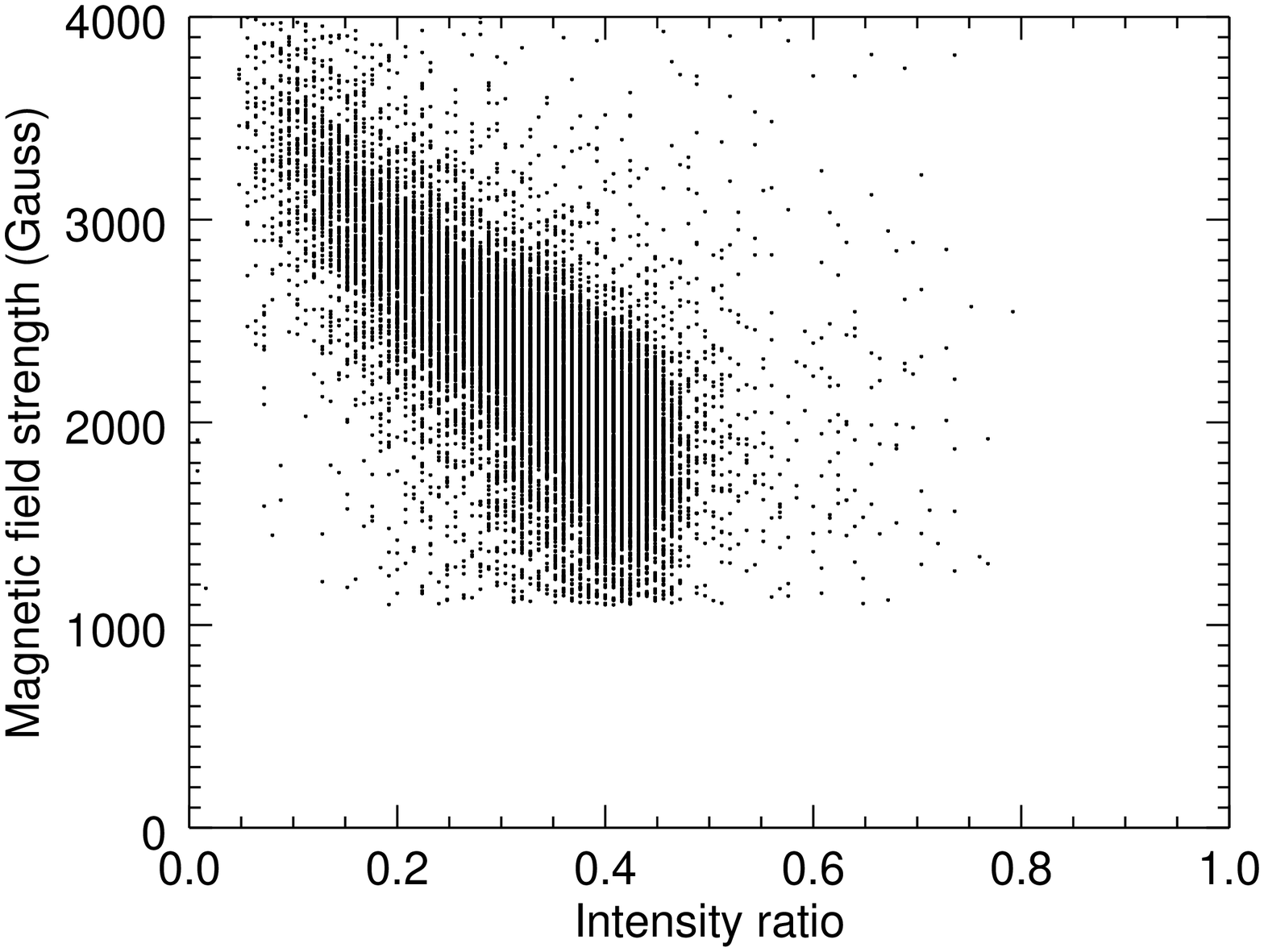}
            }
 \subfigure[HMI data]{
            \label{fig:B_vs_I_HMI}
            \includegraphics[width=0.45\textwidth]{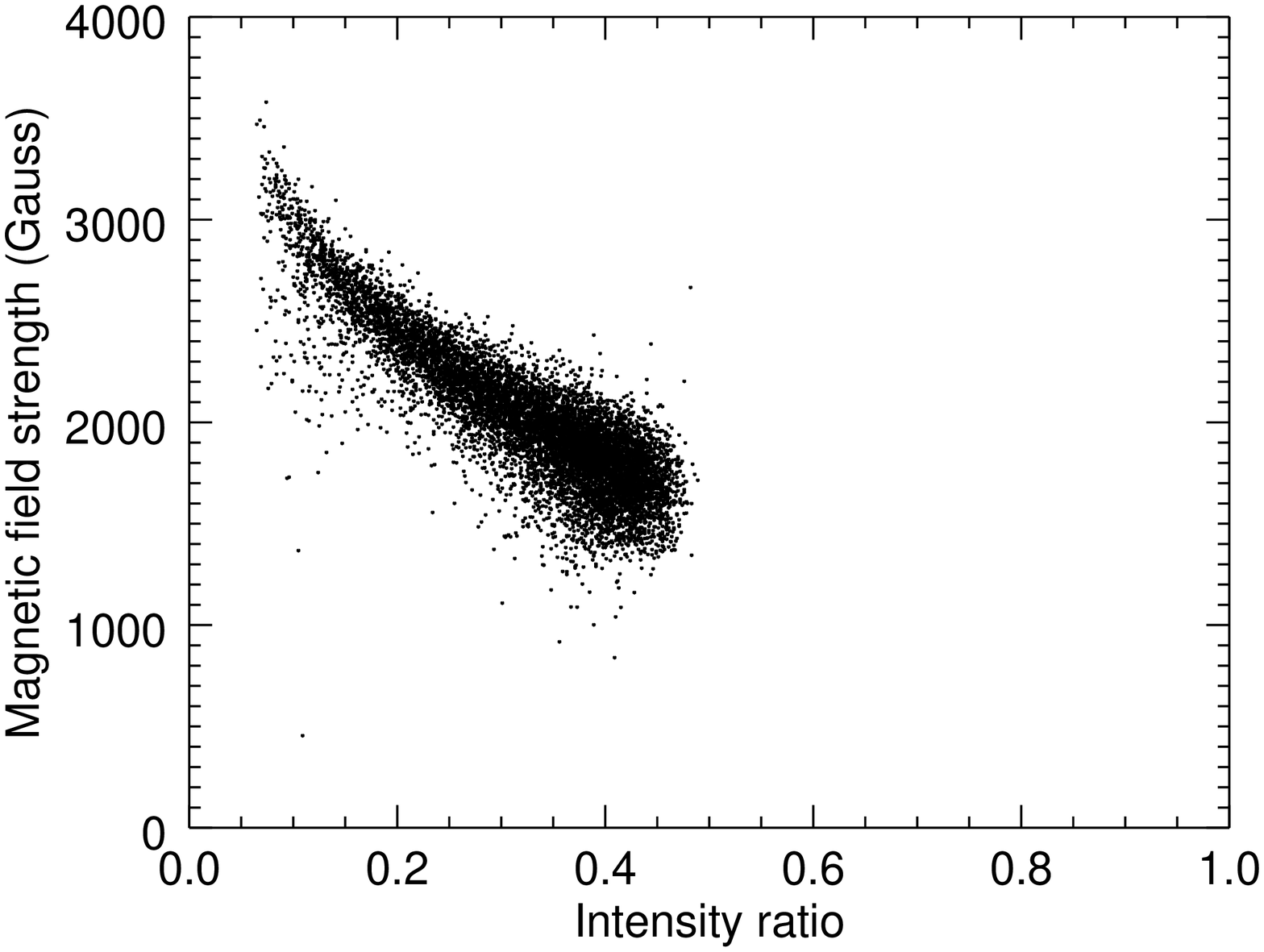}
            }
 \subfigure[BABO data]{
            \label{fig:B_vs_I_IR}
            \includegraphics[width=0.45\textwidth]{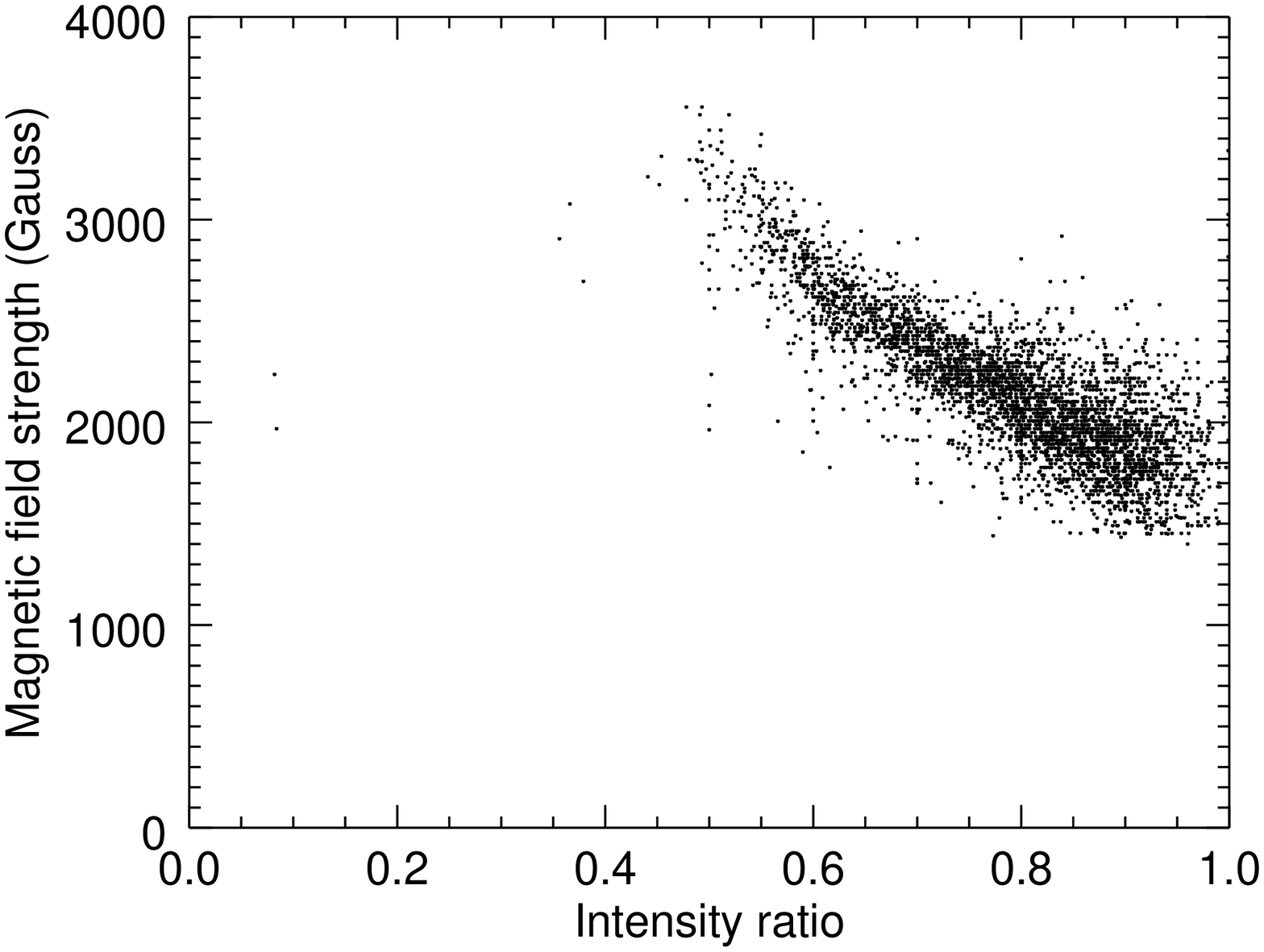}
            }
 \caption{Magnetic field strength as a function of intensity ratio for the
darkest points in sunspot umbrae for the three datasets. Intensity ratio is the
ratio between the umbral intensity and the quiet Sun intensity in the same
location.}
 \label{fig:B_vs_I}
\end{figure}

\section{Spot magnetic field and intensity relationships}\label{sect:B_vs_I}
It has long been accepted that there is a relationship between intensity at a
point on the solar surface and the magnetic flux that passes through that point.
Observations show that as the magnetic field strength increases, the intensity
decreases (\cite{1993A&A...270..494M}, \cite{1994IAUS..154..477K},
\cite{2012ApJ...745..133J}). This can be explained by the magnetic field
inhibiting convection of the plasma near the solar surface. As convection is the
primary method of heat transfer near the surface, the area with the strong
magnetic field cools and appears darker than the surrounding solar surface.

In Fig.~\ref{fig:B_vs_I}, the relationship between magnetic field strength and
intensity ratio of the darkest pixel of sunspot umbrae is shown for each
dataset.  The intensity ratio is the ratio between the umbral intensity in the
darkest pixel and the quiet Sun intensity in the same location. The quiet Sun intensity is found by interpolating pixels surrounding the sunspot. 

The MDI data shows a general decrease in magnetic field strength as the
photospheric intensity increases, in agreement with previous studies. The HMI
data show the same trend, although with a far smaller spread in values. This is
likely due to the 1.0 arc-second spatial resolution of HMI compared to the 4.0
arc-second resolution of MDI. The larger pixel size of MDI means that even when
the darkest umbral pixel is chosen, it will contain some areas with a higher
intensity and lower magnetic field. Using HMI allows this to be avoided and
although the filling factor is still not equal to one, it is higher than the
equivalent MDI measurement.

When looking at the infrared data (Fig.~\ref{fig:B_vs_I_IR}) the trend of weaker
field at higher intensities is still present, although the data occupies a
different section of intensity space. Whereas the MDI and HMI data was mostly
recorded with intensity ratios of 0.1 - 0.6, the infrared data was seen to have
intensity ratios of 0.5 - 1.0. This difference was also present in
\cite{1994IAUS..154..477K} where they measured the continuum intensity in six
sunspots on May 13th 1991 using the same infrared spectral line as used in this
article. All of their measured intensities were between 0.4 and 1.0. The 1564.8~nm spectral line was used again by \cite{2003SoPh..213...55P} to show that the magnetic field strength in NOAA active region 9885 was inversely proportional to continuum temperature, which is directly related to continuum intensity.

In the past, most authors have concentrated on only a few sunspots or active
regions and found that the field strength is proportional to intensity
throughout the spot of active region. This analysis shows that the same trend is
found when looking only at the darkest part of a sunspot umbra over the whole
spot population.

\section{Distribution of fields}\label{sect:field_dist}
Measuring the distribution of magnetic fields on the Sun as a whole can provide
insight into the mechanisms that generate the fields, particularly if studies
are over long time periods. In this section, the field distribution over time is
examined to see if there are any changes as the solar cycle progresses.

\textbf{When measuring the magnetic field with different instruments, as is the case in this article, there are a number of other effects that may change the measured values. The first of these effects is due to the different spectral line used by each instrument. HMI observes at 617.3~nm, MDI observed at 676.8~nm and BABO observes at 1564.8~nm. As such, each instrument is sensitive to a different height above the sunspot umbra depending on where the spectral line is formed. \cite{2011SoPh..271...27F} state that the HMI and MDI lines are formed at 100~km and 125~km above the solar photosphere, respectively. A study by \cite{Bruls} shows a height of formation of 18~km for the 1564.8~nm line. To determine how the height of formation affects the magnetic field measurement, we need to know how quickly the magnetic field decreases with height over a sunspot. \cite{2000ApJ...533.1035M} made simultaneous measurements of magnetic fields with two lines above sunspots and found an average magnetic field gradient of 1~Gauss~per~km. We use this value to correct the magnetic field values in this article for effects due to the height of line formation.}

\begin{figure}[ht!]
 \centering
 \includegraphics[width=0.75\textwidth]{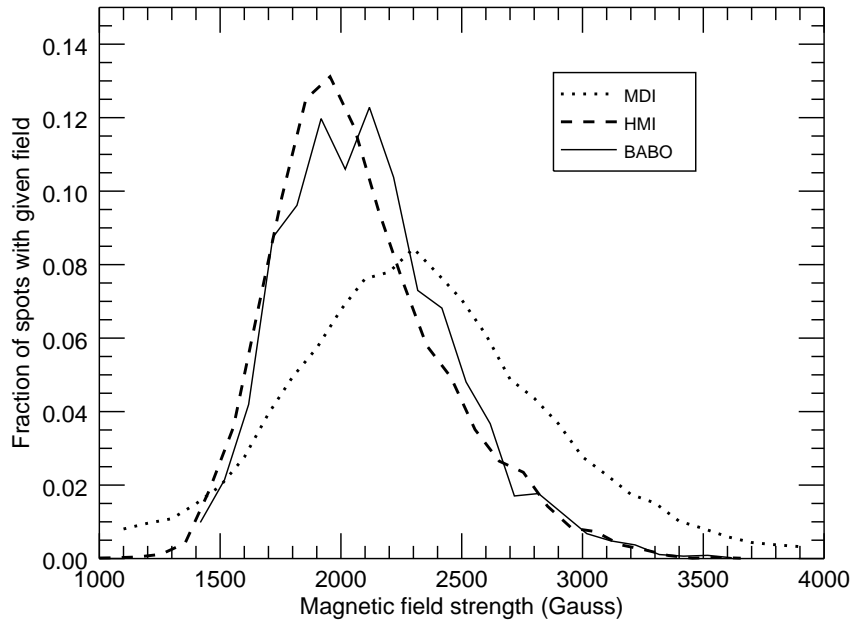}
 \caption{The magnetic field distribution of the HMI, MDI and BABO datasets.}
 \label{fig:B_hist}
\end{figure}

\begin{figure}[ht!]
 \centering
 \includegraphics[width=0.75\textwidth]{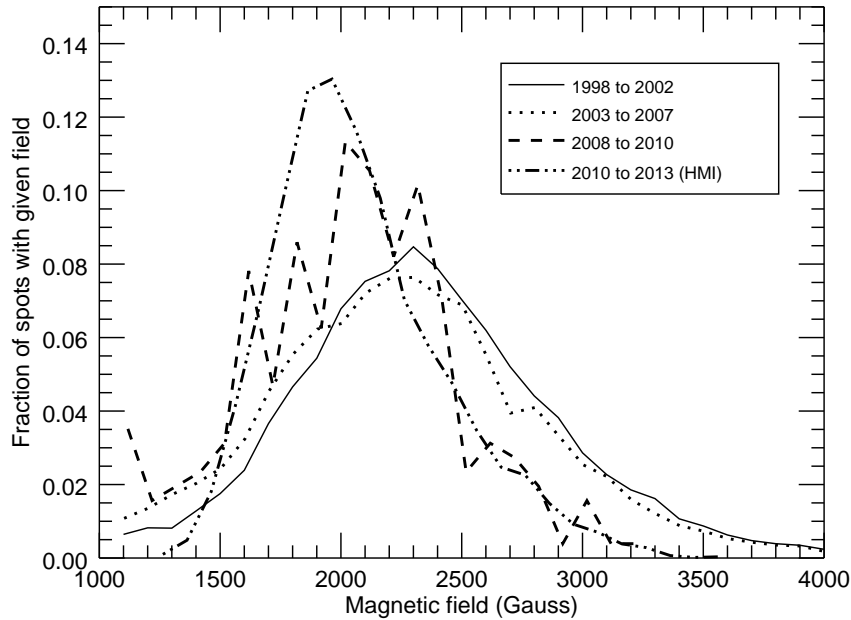}
 \caption{The magnetic field distribution measured by MDI and HMI at various
points in solar cycles 23 and 24. MDI measurements are used from 1996--2010 with
HMI measurements being used from 2010--2013. \textbf{Solar cycle 23 spans 1998 -- January 2008. Solar cycle 24 spans January 2008 -- 2013.}}
 \label{fig:B_time_hist}
\end{figure}

Fig.~\ref{fig:B_hist} shows the field distribution for the three datasets used
in this study although it should be noted that they do not cover the same time
periods. All three datasets peak around 1800--2300 Gauss with considerable
spread around their peak values. The HMI and BABO curves show a very similar
shape but with an offset of around 100~G. In Fig. 3 of
\cite{2012ApJ...757L...8L}, the
authors show the magnetic field distribution of the BABO dataset at different
times during the solar cycle. The times span the rise phase of cycle 23
(1998--2002), the maximum and fall of cycle 23 (2003--2007) and the minimum
between cycles 23 and 24 (2008-2011). Their results show that as time is
progressing, the magnetic field distribution is steadily moving to lower values
and so they also include an estimate of the field distribution for 2012--2016.
To compare with this, the field distribution of the MDI and HMI data is shown in
Fig.~\ref{fig:B_time_hist}.

The MDI and HMI data have been split into the same periods of time as the
\cite{2012ApJ...757L...8L} analysis with an exception due to the shutdown of MDI
and the activation of HMI. \textbf{The first three time periods are associated with solar cycle 23 (1998 -- January 2008) and the fourth is the onset of solar cycle 24}. The lack of MDI measurements after 2010 means that
the third time period is 2008--2010 (as opposed to 2011) and the HMI
measurements take over after that. The plot shows that for the earliest time
period, the magnetic field distribution is wider and flatter than for any of the
later periods. The 2003--2007 period is very similar to the 1998--2002 period
and between them, they cover most of solar cycle 23. There is then a change in
2008--2010 when solar minimum occurs. There are fewer spots with larger magnetic
field strengths and more with weaker field strengths. It should be noted that
there are far less spots in this time period than in any other period shown. As
cycle 24 begins (2010--2013) the sunspot umbra field distribution does not
return to what was shown between 1998--2007 but instead, continues to exhibit a
similar distribution to that of the solar minimum period albeit with more
sunspots.

This leads to the conclusion that the magnetic field distribution was
essentially constant during cycle 23, at least for the 1998--2007 period.
However the distribution shifted for the 2008--2010 period and did not recover
to what it was before. As such, the umbral fields measured with HMI in solar
cycle 24 are on average weaker and the distribution has less spread.

\section{Temporal changes}\label{sect:temporal}
\subsection{Magnetic field strengths}
Sunspots have been observed to appear on the disk as part of the solar activity
cycle and have been shown to correlate with various different indices of solar
activity (see \cite{lrsp-2010-1} and references therein). As such, this section
will examine the evolution of umbral properties over the 1996--2013 period,
covering all of solar cycle 23 as well as the rise phase of cycle 24.

\begin{figure}[ht!]
 \centering
 \includegraphics[width=0.75\textwidth]{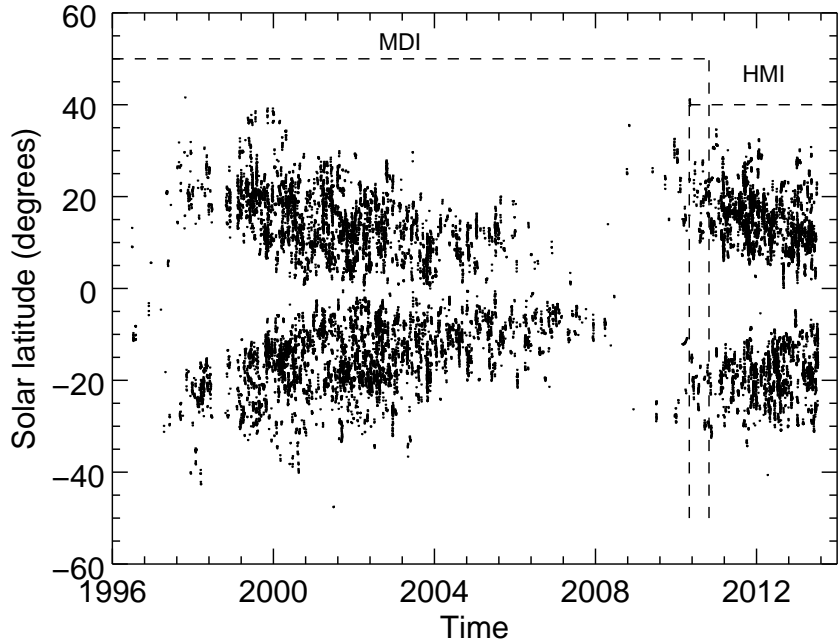}
 \caption{The latitude of sunspot appearances occupy a band on either side of
the solar equator which migrates toward the equator as a solar cycle progresses.
Both MDI and HMI sunspot detections are included and the active time of each
dataset is shown by the dashed lines.}
 \label{fig:butterfly_diagram}
\end{figure}

In Fig.~\ref{fig:butterfly_diagram}, the solar latitude of sunspots is shown as a function
of time, commonly referred to as a `butterfly diagram'. It shows the expected
pattern of sunspots appearing at preferentially higher latitudes at the
beginning of a solar cycle and migrating, on average, towards the equator as the
cycle progresses. Both MDI and HMI sunspot locations are included with the
dashed lines showing the period for which each instrument provides data.

\begin{figure}[ht!]
 \centering
 \includegraphics[width=0.75\textwidth]{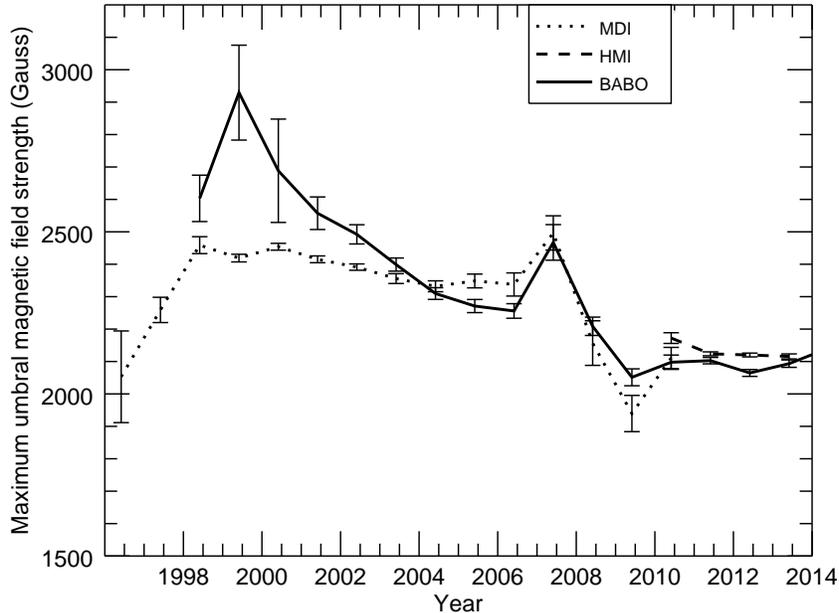}
 \caption{The annual average magnetic field in the darkest part of sunspot
umbrae for the HMI, MDI and BABO datasets. The errors bars indicate the standard
deviation of all measurements for that year.}
 \label{fig:B_time}
\end{figure}

To compare with the results in the articles mentioned in
Section~\ref{sect:intro}, the STARA sunspot catalog was used to find the darkest
pixel in each detected umbra. The corresponding magnetogram pixel was then used
to determine the magnetic field in the darkest part of the sunspot umbra. The HMI team has made Milne-Eddington
inversions of the magnetic field available and they are used whenever possible.
In order to make a more direct comparison, the HMI and MDI data has been treated
in the same manner as the BABO dataset. The average of all measurements for a
given year is plotted and the error is the standard error of the mean (equivalent to 1$\sigma$). The data
show that before around 2005, the space data did not agree well with the
infrared data measured on the ground but since 2005 there is a much better
agreement.  Fig.~\ref{fig:B_time} shows these results as well as the latest
measurements from Livingston's BABO detector.

\begin{figure}[ht!]
 \centering
 \includegraphics[width=0.75\textwidth]{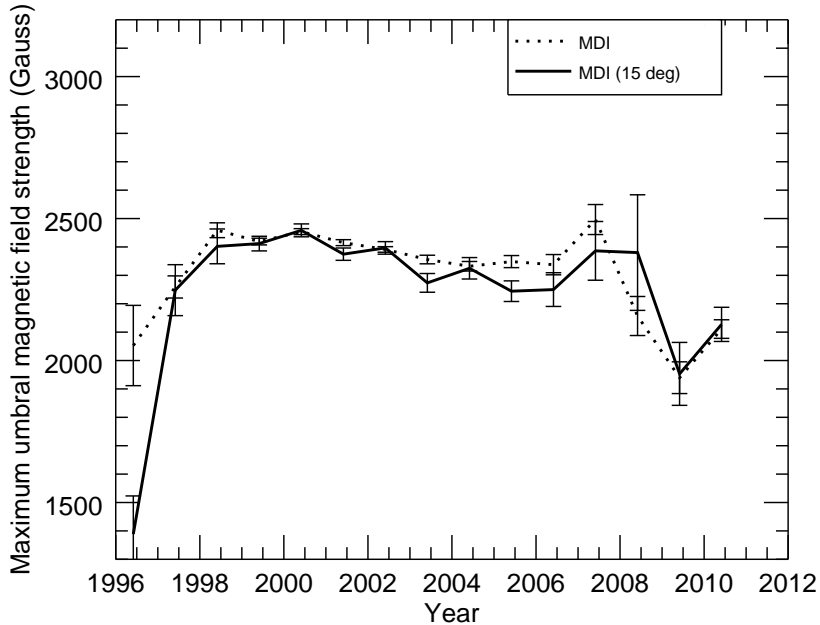}
 \caption{The same as Fig.~\ref{fig:B_time} but showing MDI data taken from a 15 degree window in longitude around the central meridian of the Sun.}
 \label{fig:MDI15deg}
\end{figure}

\textbf{The BABO dataset does not track sunspots from day to day and so the data include many instances of the same sunspot being measured. As large sunspots tend to have stronger magnetic fields and exist for longer, this could introduce a bias into the results. To test this, MDI results were compared allowing for multiple measurements of the same sunspot on consecutive days, and single measurements of sunspots. To do this, only sunspots in a longitude band of -7.5~degrees to +7.5 degrees were taken. As the Sun appears to rotate about 15~degrees per day from our point of view, this 15 degree window ensured that sunspots were only measured once. The result of this comparison is shown in Fig.~\ref{fig:MDI15deg}. The results are similar enough to justify making the comparison between BABO and MDI data using the full data set.}

From 1998--2003, there is poor agreement in the MDI and BABO datasets other than
the presence of a small downward trend after around 2000. The MDI data is around 2$\sigma$ from the BABO results over this time period. Considering the data the other way around, the BABO data is over 10$\sigma$ from the MDI results. As such, we conclude that the two sets of data do not agree with one another over the 1998--2003 time period. Also, the decrease in
umbral fields measured by BABO from 1999--2003 are eight times larger than the
corresponding decrease measured by MDI (BABO falls from 2900~G to 2250~G whereas
MDI falls from 2425~G to 2350~G). After the two datasets converge sometime in
the year 2006 there is far better agreement, and this continues with HMI data. This
raises the question of why the two datasets should be different early on and
then agree later.

The major differences between the MDI and BABO datasets are, as described in
Sect.~\ref{sect:comparison}, what is being measured and how often the
measurements are taken. The BABO data is derived from a direct measurement of
the Zeeman splitting of the Fe I 1546.8~nm spectral line whereas MDI
magnetograms involve taking points along a spectral line and fitting them to a
model. The result of this is that the individual BABO measurements are a more
accurate respresentation of the magnetic field. It measures the total field
compared to the MDI magnetogram measurement of the line-of-sight field (which is
corrected assuming a field normal to the photosphere). However, the plots in
Fig.~\ref{fig:B_time} are annual averages and so the frequency and completeness
of measurements are also factors.

To examine this, Table~\ref{table:annualmeasurements} contains the number of
umbral detections for each year from both the STARA catalog and the BABO
catalog. Also in the table are the number of days each year where at least one
umbra was successfully detected and observed. With these values it is possible
to calculate the average number of umbrae per day on days where at least one
umbra was present.

\begin{table}
\begin{center}
\begin{tabular}{ | c | c | c | c | c | c | c | }
  \hline
  \multirow{2}{*}{Year} & \multicolumn{3}{|c|}{MDI / HMI} &
\multicolumn{3}{|c|}{BABO} \\
   & \# & Days & \# per day & \# & Days & \# per day \\ \hline
  1996 & 35 & 13 & 2.69 & - & - & - \\ \hline
  1997 & 178 & 67 & 2.66 & - & - & - \\ \hline
  1998 & 570 & 159 & 3.58 & 10 & 3 & 3.33 \\ \hline
  1999 & 2265 & 287 & 7.89 & 6 & 2 & 3.00 \\ \hline
  2000 & 3018 & 307 & 9.83 & 4 & 2 & 2.00 \\ \hline
  2001 & 3071 & 323 & 9.51 & 56 & 14 & 4.00 \\ \hline
  2002 & 3419 & 317 & 10.79 & 105 & 14 & 7.50 \\ \hline
  2003 & 1896 & 275 & 6.89 & 302 & 29 & 10.41 \\ \hline
  2004 & 1143 & 230 & 4.97 & 249 & 34 & 7.32 \\ \hline
  2005 & 819 & 282 & 4.01 & 282 & 37 & 7.62 \\ \hline
  2006 & 356 & 150 & 2.37 & 179 & 29 & 6.17 \\ \hline
  2007 & 154 & 84 & 1.83 & 60 & 18 & 3.33 \\ \hline
  2008 & 41 & 18 & 2.28 & 61 & 10 & 6.10 \\ \hline
  2009 & 54 & 21 & 2.57 & 79 & 19 & 4.16 \\ \hline
  2010 & 564 & 176 & 3.20 & 223 & 47 & 4.74 \\ \hline
  2011 & 3500 & 340 & 10.29 & 1059 & 56 & 18.91 \\ \hline
  2012 & 3554 & 360 & 9.87 & 911 & 66 & 13.80 \\ \hline
\end{tabular}
\caption{Number of umbrae measured}
\label{table:annualmeasurements}
\end{center}
\end{table}

\begin{figure}[ht!]
 \centering
 \includegraphics[width=0.75\textwidth]{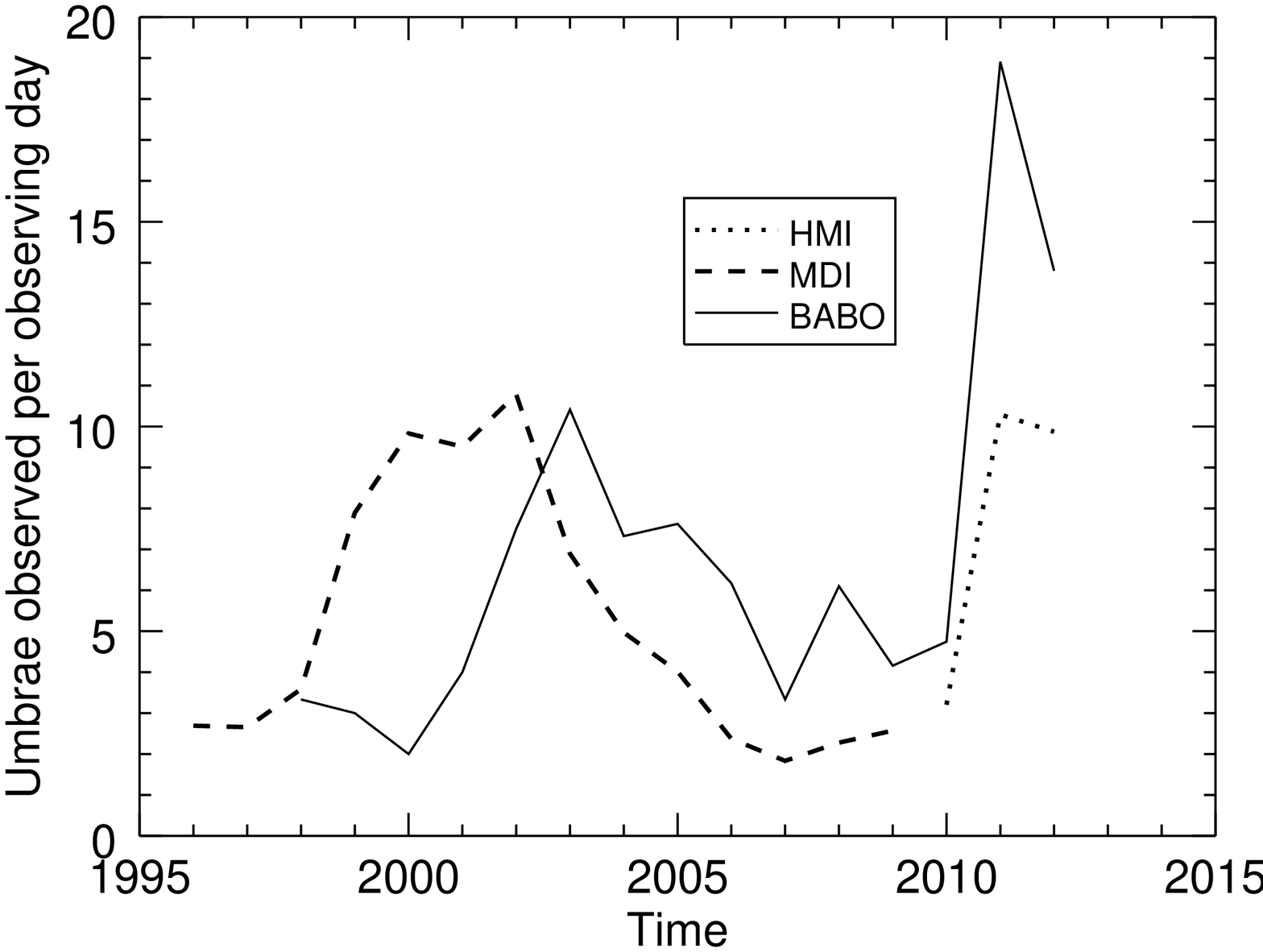}
 \caption{The average number of umbrae observed per observing day. Only days
where at least one umbra was observed are included. The time of disagreement between BABO and MDI in this figure is 1998--2003. The large increase in umbrae observed with BABO from 2010--2013 are due to more careful observations of large numbers of small sunspots compared to previous times.}
 \label{fig:average_umbrae_observed}
\end{figure}

It should be noted here that the number of days each year that yielded a
successful umbral measurement is not the same as the number of observing days
per year, particularly during times of low solar activity. This would be an
unfair metric to use when comparing the datasets as MDI and HMI are taking
measurements many times per day whereas the BABO dataset relies on the
availability of both the telescope and the observer to make the observation. In
addition to this, the fact that BABO was used less often than MDI or HMI does
not mean that it is not able to measure a sample that is representative of the
full sunspot population. That is what will be addressed in this section.

Table~\ref{table:annualmeasurements} shows
that the number of days that BABO observed sunspot umbrae was very low from 1998
until the end of 2000 and so this could explain why the datasets disagree over
this time period. From the beginning of 2001, more frequent observations were
taken with BABO at the McMath-Pierce Solar Telescope starting with 14 days in
the year 2000 and even reaching 66 separate successful observation days in 2012
(equivalent to a 5 or 6 day observation run each month although some days are
lost due to weather). 

As both datasets did not have the same observing cadence, it is beneficial
to look at the number of umbrae observed per successful observing day which is
shown in Fig.~\ref{fig:average_umbrae_observed}. This is calculated by taking
the total number of umbrae detected in a given year and dividing by the number
of days when at least one umbra was detected. It should eliminate effects from
the different observation cadences.

The MDI and HMI data show a cyclic trend in the average number of umbrae
observed per day and even show the double peaked structure of solar cycle 23
with highs in 2000 and 2002 (when binned annually). The BABO data shows a noisy
cyclic trend with a rise from 2000--2003 then falling to a minimum in
2007--2009. This leads to the conclusion that the STARA sunspot umbra detections from the MDI dataset can be thought of as a good representation of the global sunspot
population due to the fact that the number of sunspot umbrae detected correlates
with the sunspot number from other sources. The BABO data appears to follow the sunspot number from 2003 onwards but they are not in agreement before that time. In 2011 and 2012, very large numbers of umbrae were detected per day in the BABO set, far more than at any point during the previous cycle. This is strange as the sunspot number in 2011 and 2012 has been comparable to that in 2003. It is also shown that the HMI number of umbrae observed per day in 2011 is comparable to the MDI value in 2002 when the
sunspot number is different. This is likely because the increased spatial
resolution of HMI allows smaller umbrae to be detected and so more are observed
on average. The BABO number of umbrae observed per day are systematically higher than the automated MDI and HMI values. This is a result of the automated nature of those detections. To ensure the lowest number of false sunspot detections possible, very small sunspots are not included in the detections. This is not the case with the manual BABO dataset and so more umbrae are observed per day. Due to the method of measurement, the extra spots in the BABO data are considered very small.

The data in Table~\ref{table:annualmeasurements} along with Fig.~\ref{fig:average_umbrae_observed} seem to
show an issue in the sampling of the sunspot population with the BABO dataset.
As mentioned previously, the BABO detector can provide more accurate magnetic
field strength measurements than MDI or HMI. However, the apparent change in the
BABO population sample over time makes extrapolating long-term trends difficult.
By examining a property that is independent of the magnetic field, in this case
the observation frequency and the number of umbrae detected, it appears that the
number of umbrae measured in the BABO dataset is not correlated with the sunspot
number until sometime in the 2003--2005 period after which a trend can be seen.
This implies that the BABO dataset is not a representative sample of the full
sunspot umbrae population before 2003 and provides an explanation for the
difference between the umbral magnetic fields measured by BABO and by MDI and
HMI. This is reinforced by Fig.~\ref{fig:B_time} which shows that from 2005
onwards, the annual umbral magnetic field strengths measured by all instruments
are in agreement.

\begin{figure}[ht!]
 \centering
 \subfigure[Northern hemisphere]{
            \label{fig:B_North}
            \includegraphics[width=0.65\textwidth]{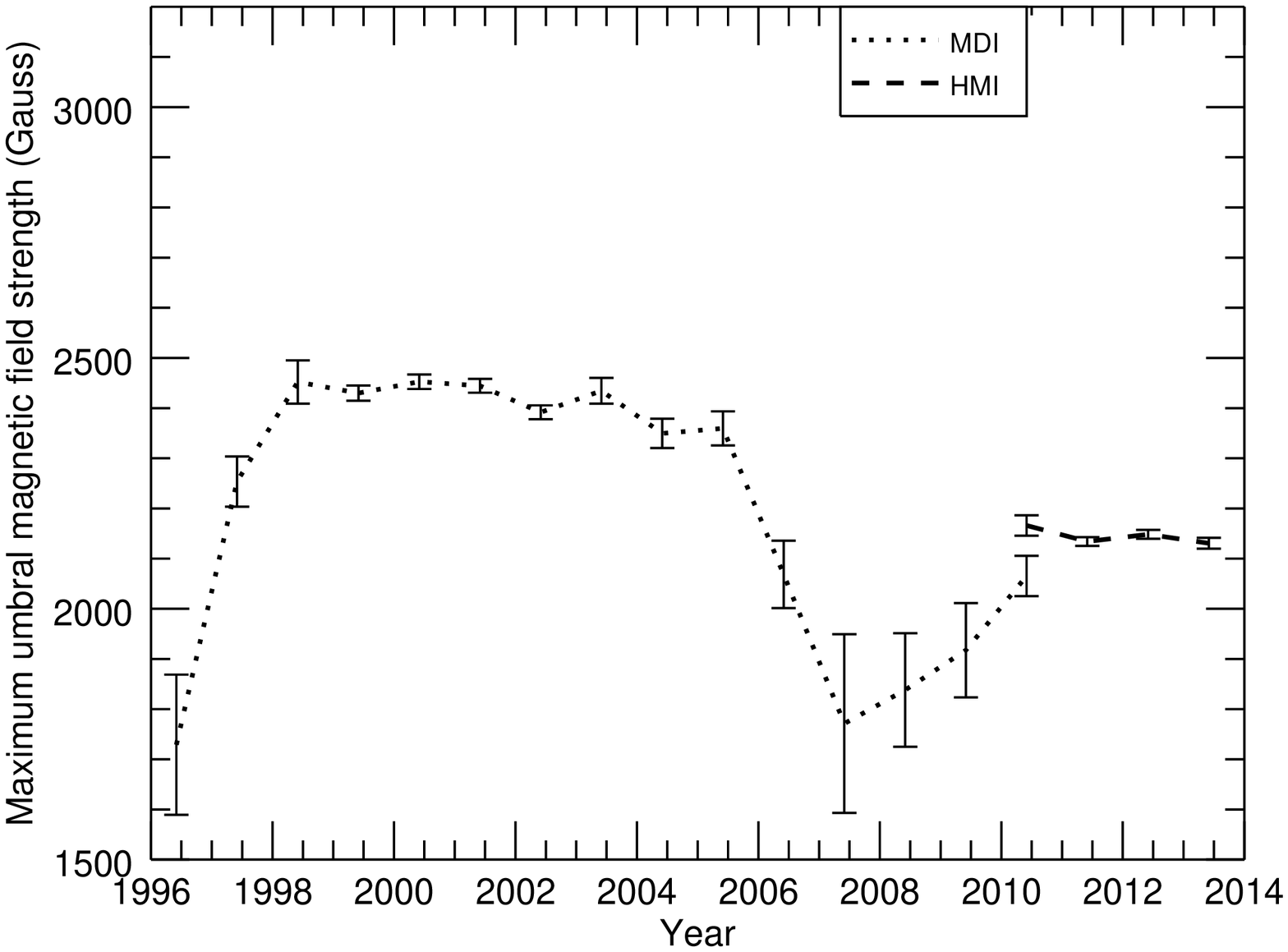}
            }
 \subfigure[Southern hemisphere]{
            \label{fig:B_South}
            \includegraphics[width=0.65\textwidth]{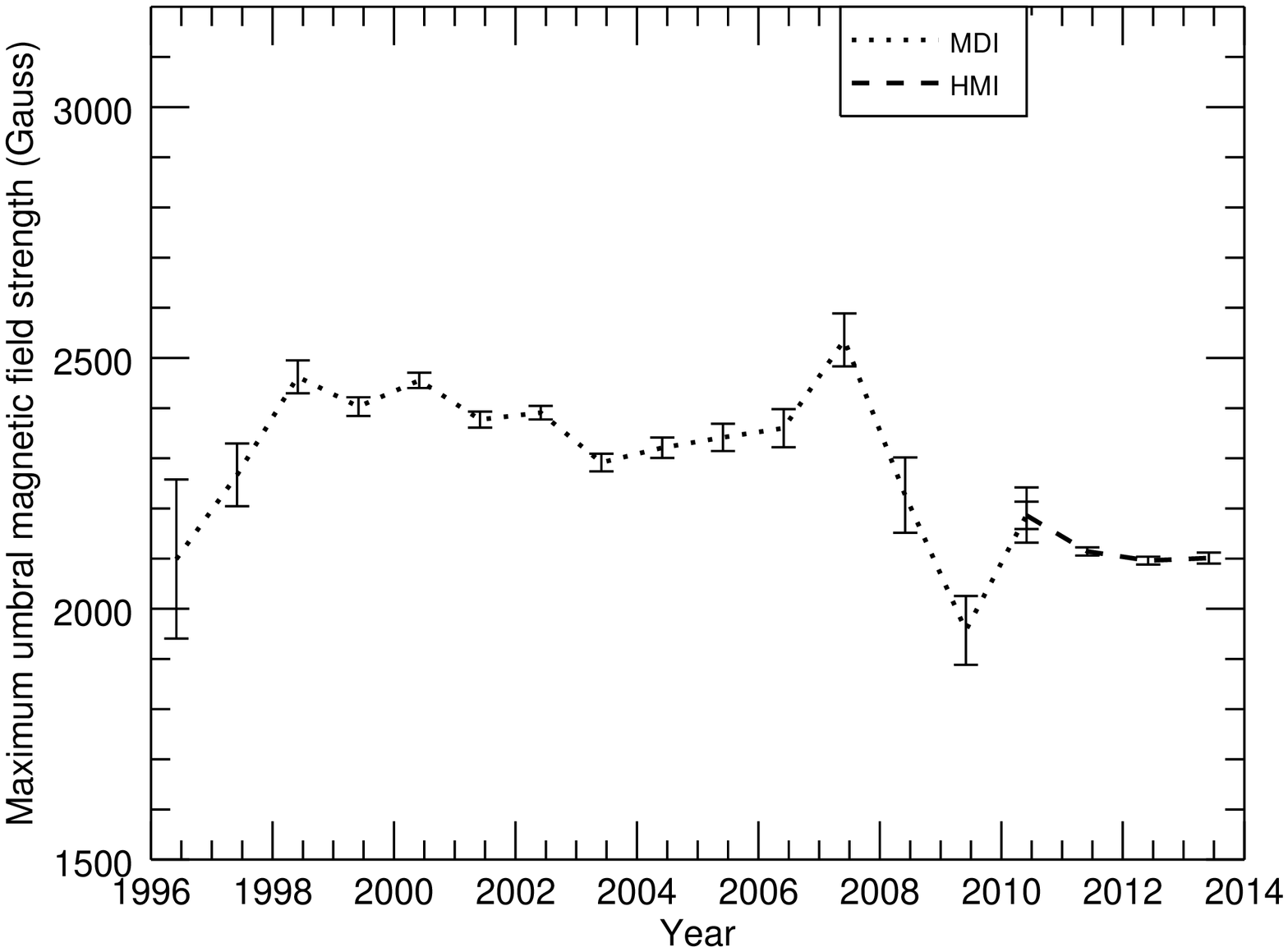}
            }
 \caption{MDI and HMI data treated in the same manner as Fig.~\ref{fig:B_time} but split by hemisphere. BABO data is not included as position of the sunspots is not readily available in the BABO catalog.}
 \label{fig:B_NS}
\end{figure}

\textbf{It has been previously shown that the solar activity levels in the North and South hemispheres of the Sun are out of sync (see \cite{1986ApJ...307..389H} and \cite{2003SoPh..214...23D}), particularly during polar reversals. We show the magnetic field values of MDI and HMI in Fig.~\ref{fig:B_NS} but this time, split by hemisphere. This shows that the drop in sunspot magnetic fields during the minimum between cycles 23 and 24 was larger in the Northern hemisphere than in the Southern. In addition, the lowest value of the maximum field in the Northern hemisphere occured 6--12 months before the equivalent value in the Southern hemisphere indicating a delay as reported by previous studies. Including the BABO data in these plots was not possible as the location of the sunspot on the disk was not listed in the catalog.}

\subsection{Umbral intensities}
\begin{figure}[ht!]
 \centering
 \includegraphics[width=0.75\textwidth]{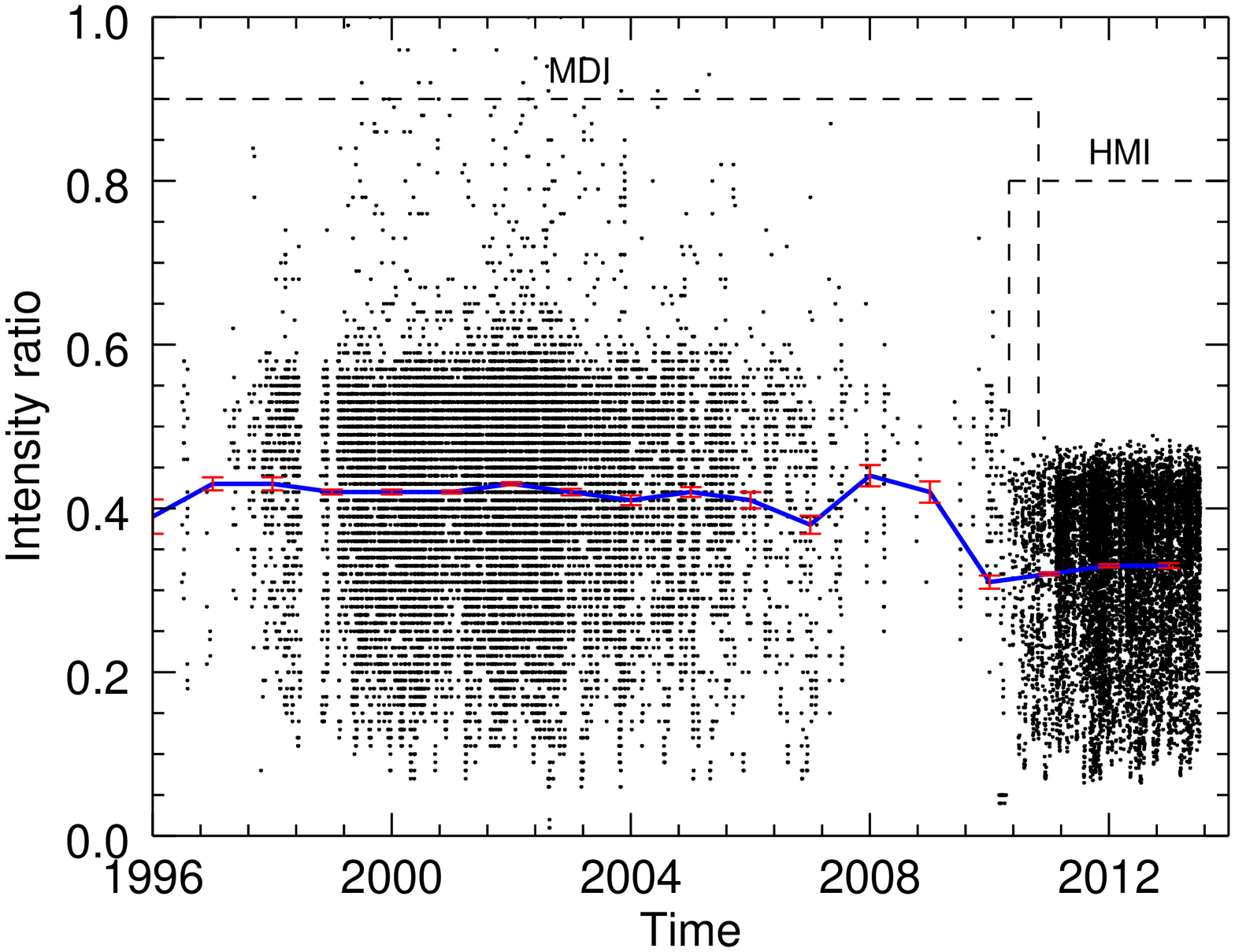}
 \caption{The ratio between darkest umbral intensity and quiet Sun intensity for
all sunspots seen in the MDI and HMI STARA catalogs. An annual mean is also
shown in blue.}
 \label{fig:Intensity_vs_time}
\end{figure}

Besides examining the magnetic field strengths, it is also possible to look at
how the intensity of sunspot umbrae change over time. Fig.~\ref{fig:B_vs_I} has
already shown a correlation between umbral intensity and magnetic field and so
we would expect this to be reflected in the long term measurements of
intensities. The intensities of sunspots and sunspot umbrae
have been the subject of many studies including \cite{1984SoPh...90...17A},
\cite{1986ApJ...306..284M}, \cite{2004ApJ...603..348N},
\cite{2006ApJ...649L..45P}, \cite{2007A&A...465..291M},
\cite{2012ApJ...757L...8L} and \cite{2012rezaei} (see \cite{2013JPhCS.440a2038N}
for a comparison of some of these studies). Even more recently,
\cite{2013ApJ...771L..22D} have used images from the San Fernando Observatory to
examine the intensities of spots and umbrae. The variation in the sample size
used in these studies is detailed in Table~\ref{table:intensities} and ranges
from 22 umbrae to almost 27000. The intensities of all umbrae in the STARA
catalog along with an annual mean are shown in
Fig.~\ref{fig:Intensity_vs_time}.

\begin{table}
\begin{center}
\begin{tabular}{ | c | c | c | c | c | c | c | }
  \hline
  Study & \# & Instrument \\ \hline
  Albregtsen et. al. (84) & 22 & Tower telescope, Oslo \\ \hline
  Norton \& Gilman (04) & 650 & MDI \\ \hline
  Mathew et. al. (07) & 160 & MDI \\ \hline
  Livingston et. al. (12) & 2700 & BABO, McMath-Pierce \\ \hline
  Rezaei et. al. (12) & 183 & Tenerife Infrared Polarimeter \\ \hline
  de Toma et. al. (13) & 22765 & CFDT1, San Fernando Obs. \\ \hline
  This article & 26921 & MDI and HMI \\ \hline
\end{tabular}
\caption{Studies measuring sunspot intensities.}
\label{table:intensities}
\end{center}
\end{table}

The plot shows that there is a fundamental difference in the intensity ratio
numbers between MDI and HMI. The MDI data is mostly in the 0.60 to 0.05 range
whereas HMI data is found in the 0.50 to 0.05 range. In addition to this, the
running mean is fairly flat from 1996--2010 which contains all of the MDI data
but after switching to HMI observations, the running mean drops from 0.43 to
0.33 almost immediately. This indicates that there is either an issue with the
method used to measure the intensity ratio (which is the same for both
instruments) or that there is a difference in the data. The latter explanation
is most likely as these instruments observe in different spectral lines (MDI in
676.8~nm and HMI in 617.3~nm). The calculated intensity is a linear combination
of filtergrams taken at various points in the spectral line profile and there is
no reason why the combination should be the same for two different spectral
lines. However if each dataset is examined independently then the intensity
ratio is, on average, mostly constant over time.

This would appear to contradict the earlier results given in this article. We have shown in Fig~\ref{fig:B_vs_I} that the umbral intensity is related to the umbral magnetic field strength. Then, in Fig~\ref{fig:B_time}, we show that the average magnetic field in the darkest part of sunspot umbrae has decreased by around 325~G from 1998--2010 as measured by MDI. If both these results are true, it seems counterintuitive that the umbral intensity from 1998--2010 is almost constant. This can be explained by considering the range of possible magnetic field measurements for a single umbral intensity. The mean umbral intensity from 1998--2010 is around 0.43 from Fig~\ref{fig:Intensity_vs_time}. Looking back at Fig~\ref{fig:B_vs_I}, we can see that for an umbral intensity ratio of 0.43, the range of possible magnetic field strengths measured by MDI is over 1000~G, and with the improved HMI instrument that range drops to around 650~G. This leads to the conclusion that the umbral magnetic field could change by the measured 375~G without a significant change in umbral intensity. And so, the lack of changing intensity shown here is consistent with the results shown earlier in this article. Physically, this intensity change in a sunspot with a given maximum magnetic field may be related to the twist of the magnetic field geometry as discussed in \cite{1993A&A...270..494M}.

To allow the MDI and HMI intensity data to be compared with other data, the two most `complete' studies from
Table~\ref{table:intensities} are used, with complete meaning the most
observations over the longest timescale. This leads to the
\cite{2012ApJ...757L...8L} and \cite{2013ApJ...771L..22D} studies.

Fig. 2 in the \cite{2012ApJ...757L...8L} article shows the raw intensity ratio
data along with annual averages measured with the BABO detector. It shows a
clear upward trend over the 1998--2012 range plotted and also shows some sunspot
umbrae measured in 2011 and 2012 with intensities well above 0.95. The upward
trend starts at a value of around 0.60 in 1998 and rises to around 0.85 in 2012.
This is in agreement with the magnetic field trend also reported in that article
as an increase in sunspot umbral intensity would imply a decrease in umbral
magnetic fields. However, earlier in this article a possible explanation for
differences in magnetic field strength between datasets was low sampling rates
in the BABO dataset before 2005. If this caveat is applied here there is still a
small upward trend in the BABO data although it is now from around 0.70 in 2005
to 0.85 in 2012. This range of 0.15 is approximately equivalent to the range
seen in the MDI running mean intensity in Fig.~\ref{fig:Intensity_vs_time}. Note
that it is the trends that are more important here as the different spectral
lines used make direct intensity comparisons difficult.

Fig. 2 in the \cite{2013ApJ...771L..22D} article also shows sunspot umbral
intensities but this time using the Cartesian Full Disk Telescope 1 (CFDT1) at a
wavelength of 672.3~nm at San Fernando Observatory. This wavelength is different
to those probed by MDI, HMI and BABO. A primary advantage of this dataset is
that it covers the period of 1986--2012, substantially longer than any dataset used in this article. The umbral intensity data given spans a very large range, from 0.90 down to 0.15 but the trend is similar to that seen in the MDI and HMI data. Very little variation is seen in the average umbral intensity over time.

\textbf{Although ground observations contain more scattered light from outside the instrument due to the atmosphere, these results show that even comparing ground-based telescopes does not provide the same result. A similar study by \cite{2013SoPh..tmp..263S} shows the same result of sunspot umbral intensities being almost constant using observations from HINODE. All of the visible light studies agree with the visible light MDI and HMI results presented in the article. It would then appear that there is a possible difference between instruments measuring visible light and infra-red light.}

\textbf{An article by J.E. Harvey et. al. (2012) shows that instrumental stray light varies as $1/\lambda^2$ and using the values for MDI and BABO, this means that the stray light within the instrument is around 5.3 times bigger for MDI than for BABO. This equation assumes reflective optics and the transmissive optics in MDI will make this value worse. In addition to this, the sunspot contrast is greater in visible light than in infra-red light. This means that observations in visible light are more susceptible to scatter from outside the umbra than infra-red observations. In all of the MDI and HMI intensity measurements, there is a large amount of scatter that could be masking a trend in intensity. As such, it is possible that the BABO observations are seeing a trend that either cannot be seen in visible light due to scattered light effects, or that only exists in infra-red emission. To check further, it would be advantageous to have similar infra-red observations from another observatory.}

\section{Summary and conclusions}\label{sect:conclusions}
The magnetic fields and umbral intensities of sunspots have been analysed using
both ground based and space based instrumentation with comparisons drawn between
the datasets. The MDI and HMI intensity data was processed using an automated
sunspot detection algorithm to build a self consistent catalog containing
sunspot magnetic information from magnetograms taken by the same instruments.

The well established magnetic field - intensity relationship was shown to hold
for the MDI, HMI  and BABO datasets with the HMI data providing a much tighter
correlation due to the increased spatial resolution and lower noise level. The
distribution of magnetic fields within each dataset was plotted and showed that
the HMI and BABO distributions were similar, with the MDI distribution
exhibiting a wider, flatter profile.

The temporal evolution of sunspot umbral fields showed a decrease in the mean umbral field strength although the decrease was not as large as that previously reported by
\cite{2012ApJ...757L...8L}. When the annual values were compared between the
space instruments and BABO, it was found that the agreement is far better after
2005 than before. By looking at the observation cadence and the number of umbrae
detected per observation, a possible explanation was found concerning the number
of observations in the BABO dataset. As the umbral detection rate is expected to
correlate with the sunspot number, and this is only the case with the BABO
dataset after sometime between 2003 and 2005, we conclude that the BABO data
before that time is not representative of the full sunspot population.

A similar comparison was then made for the umbral intensities in sunspots and showed that, on average, the umbral intensities of sunspots did not vary over time. This was shown to be in agreement with a magnetic field strength that decreases with time as the range of magnetic field values measured for spots with a given intensity was larger than the decrease in the magnetic field strength over time.

\textbf{A recent study by \cite{2014SoPh..289.1143T} showed that some parameters (including the average magnetic flux in sunspots) exhibited a bimodal distribution that is not seen in the data used in this article. This is because their method of automated identification measures both pores and sunspots in HMI data. The bimodal distribution they found was one mode below an area of 20~MSH (mostly populated by pores) which comprised of at least 65\% of their data, and the other above an area of 100~MSH (exclusively populated by sunspots). As most of the data in our study comes from MDI, which has a lower spatial resolution than HMI, we are not able to measure most of these very small features. It is possible that we would see the bimodal distribution by focussing solely on HMI data. As such, we did not expect to see the same bimodal distribution observed by \cite{2014SoPh..289.1143T} in this study.}

\section*{}
We would like to thank our referee for their valuable comments that have improved the content of this study. SOHO is a project of international cooperation between ESA and NASA. The HMI
data used are courtesy of NASA/SDO and the HMI science team. The McMath-Pierce
Solar Telescope is operated by the National Solar Observatory, which is managed
by the Association of Universities for Research in Astronomy under contract with the National
Science Foundation.

\bibliographystyle{spr-mp-sola}
\bibliography{bibliography}

\end{document}